\documentclass{article}
\usepackage[utf8]{inputenc}

\usepackage{mathrsfs}
\usepackage{amsfonts}
\usepackage{cite}
\usepackage{epsf,epsfig,amsfonts,amsgen,amsmath,amstext,amsbsy,amsopn,amsthm}
\usepackage{hyperref}

\theoremstyle{definition}

\begin{document}

\newcommand{\tabincell}[2]{\begin{array}{@{}#1{}}#2\end{array}}

\title{\bf {\ Optimal Quaternary Hermitian  LCD  codes}}
\date{2020.10.08 (Manuscript )}
\author{Liangdong Lu$^{a,\dag}$, Xiuzhen Zhan$^{a}$, Sen Yang$^{a}$, Hao Cao$^{b,\dag\dag}$\\
a. College of  Science,  Air Force Engineering University,  Xi'an,
 Shaanxi,\\ 710051, China, (email:
$^{\dag}$ kelinglv@163.com, )\\
  b.School of Information and Network Engineering, Anhui Science \\and Technology
  University,
  Chuzhou, 233100, China
School of Mathematical Science, \\Huaibei Normal University,
Huaibei, 235000, China (email:$^{\dag\dag}$13655505689@163.com) \\}
\maketitle

\begin{abstract}
Linear complementary dual (LCD) codes, which is a class of linear
codes introduced by Massey,  have been extensively studied in
literature recently. It has been shown that LCD codes can help to
improve the security of the information processed by sensitive
devices, especially against so-called side-channel attacks (SCA) and
fault invasive attacks. In this paper, Tables are presented of good
quaternary Hermitian LCD codes and there are used in the
construction of puncturing, extending, shortening and combination
codes. Results including tables 3 of the best-known quaternary
Hermitian LCD codes of any length $ n \leq 25$ with corresponding
dimension $k$ are presented. In addition,  Many of these quaternary
Hermitian LCD codes given in this paper are optimal which are
saturating the lower or upper bound of Grassl's codetable in
\cite{Grassl}  and some of them are nearly optimal.

\end{abstract}
{\bf Keyword:} quaternary Hermitian, Linear complementary dual,
linear code, optimal.

\section{Introduction}
\noindent

Let $q$ be a power of a prime p, $F_{q}$ be the finite field with
$q$ elements, and $F^{n}_{q}$ be the $n$-dimensional vector space
over $F_{q}$. A $q$-ary $[n,k,d]_{q}$ linear code over $F_{q}$ is a
$k$-dimensional subspace of $F^{n}_{q}$ with Hamming distance $d$.
For a given  $[n,k]_{q}$ linear code, the code
$\mathcal{C}$$^{\perp}=\{x|x\cdot c=0, c\in \mathcal{C}\}$ is called
the dual code of $\mathcal{C}$\cite{MacWilliams,Huffman}. A $q$-ary
linear code $\mathcal{C}$ is called a linear complementary dual
(LCD) code if it meets its dual trivially, that is
$\mathcal{C}\cap\mathcal{C}^{\perp} =\{\mathbf{0}\}$, which was introduced by Massey
\cite{Massey1964,Massey1992}. In addition to their applications in
data storage, communications systems, and consumer electronics, LCD
codes have been employed in cryptography and quantum error
correcting recently. Carlet and Guilley in \cite{Carlet2016} shown
that LCD codes play an important role in armoring implementations
against side-channel attacks, and presented several constructions of
LCD codes.

 In \cite{Lu2014}, according to finite geometry theory,
Lu et.al proposed the {\it radical codes} $\mathcal{R(C)}$ of
$\mathcal{C}$ and $\mathcal{C}^{\perp}$ which is
$\mathcal{R(C)}=\mathcal{C}\cap \mathcal{C}^{\perp}$. If
$\mathcal{R(C)}=\mathcal{C}\cap \mathcal{C}^{\perp}=\{0\}$, then
$\mathcal{C}$ is called to be a zero radical code, which is the same
as the LCD code presented in \cite{Massey1992}. Using these obtained
zero radical codes, they construct families of Maximal entanglement
entanglement-assisted quantum error-correcting codes which can help
to engineer more reliable quantum communication schemes and quantum
computers. Furthermore, constructions of Hermitian zero radical  BCH
codes
 were discussed  \cite{lu2015}, which are also called
reversible codes in \cite{Massey1964} or LCD cyclic codes in
\cite{Ding2016xiv}. C. G\"{u}neri et.al studied quasi-cyclic LCD
codes and introduced Hermitian LCD
 codes \cite{uneri2016}. Moreover, for Euclidean case,  the question that when cyclic
 codes are LCD codes is answered affirmatively by Yang and Massey in
\cite{Yang1994}. Ding {\it et.al} investigate LCD cyclic code
\cite{Ding2016xiv} in which  several families of LCD cyclic codes
are constructed. It is shown that some LCD cyclic codes are optimal
linear codes or have the best possible parameters for cyclic codes.
Currently, many works have focused on the construction of LCD codes
with good parameters, see
\cite{Hou,Lina,Shixin,Galvez,Araya1,Araya2,Fu}.

Recently, Carlet, Mesnager, Tang, Qi, Pellikaan in \cite{Carlet2018}
shown that any $[n, k, d]$-linear code over $F_{q^{2}}$ is
equivalent to  an $[n, k, d]$- linear Hermitian LCD code over $F_{q^{2}}$ for $q > 2$.
Araya, Harada and Saito in \cite{Araya3} give
some conditions on the nonexistence of quaternary Hermitian linear
complementary dual codes with large minimum weights. Inspired by
these works and extending our previous work in \cite{Lu2014},   we
study constructions of linear Hermitian LCD code over $F_{4}$. Then
some families of linear Hermitian LCD codes with good parameters are
constructed  from the known optimal codes by puncturing, extending,
shortening and combination method. Compared with the tables of best
known linear codes (referred to as the {\it Database} later)
maintained by Markus Grassl at $http://$ www.codetables.de, some of
our codes presented in this paper are saturating the lower bound of
Grassl's codetable.

In this paper, Optimal quaternary Hermitian LCD codes $[18,7,9]$,
$[19,7,9]$ and $[20,7,10]$ are given, which improve the minimal
distance of the codes in\cite{Lu2014,lai3}. According to
classification codes in \cite{Bouyukliev}, there exist some optimal
quaternary Hermitian  LCD codes $[15,4,8]$, $[16,4,9]$, $[17,4,10]$,
$[19,4,13]$, $[23,5,14]$  and $[15,6,7]$. According to
\cite{Grassl}, the following quaternary Hermitian  LCD codes we give
in this section are also optimal linear codes: $[n,k, d]$ for
$21\leq n\leq 25$ and $16\leq k\leq 18$; $[n-i,15-i, d]$ for $21\leq
n\leq 24$ and $0\leq i\leq 2$;  $[23, 4, 15]$, $[24, 5, 15]$, $[21,
6, 12]$, $[22, 6, 12]$, $[23, 6, 13]$, $[24, 6, 14]$, $[25, 8, 12]$
and $[20, 7, 10]$.

 This paper is organized
as follows. In Section 2 we provide some required basic knowledge on
Hermitian  LCD codes. We derive constructions of Hermitian LCD codes
in Section 3. In Section 4, we discuss Hermitian LCD  codes with
good parameters.

\section{ Preliminary}
\noindent

In this section,  we introduce some  basic concepts on quaternary
linear codes. Let $\mathbf{F}_{4}=\{0,1,\omega,\varpi\}$  be the
Galois field with four elements with
$\varpi=1+\omega=\omega^{2},\omega^{3}=1$, and the conjugation is
defined by $\bar x=x^{2}$ for $x\in\mathbf{F}_{4}$. Denote the
$n$-dimensional space over $\mathbf{F}_{4}$ by $\mathbf{F}_{4}^{n}$,
we call a $k$-dimensional subspace $\mathcal{C}$ of
$\mathbf{F}_{4}^{n}$ as an $k$-dimensional linear code of length $n$
and denote it as $\mathcal{C}$ $=[n,k]$. A matrix $G$ whose rows
form a basis of $\mathcal{C}$ is called a generator matrix of
$\mathcal{C}$. If the minimum distance of $\mathcal{C}$ is $d$, then
$\mathcal{C}$ can be denoted as $\mathcal{C}$ $=[n,k,d]$. A code
 $\mathcal{C}$$=[n,k,d]$ is an {\it optimal} code if there is no
$[n,k,d+1]$ code. If  $d$ is the largest
 value present known that there  exists  an $[n,k,d]$, then
$\mathcal{C}$$=[n,k,d]$ is called a {\it best known }  code.
 Denote $d_{l}(n,k) = max \{d|$  an $[n,k,d]$ LCD code$\}$.
 If an $\mathcal{C}=[n, k, d_{l}(n,k)]$ LCD
code is saturating the lower or upper bound of Grassl's
codetable\cite{Grassl}, we call $\mathcal{C}$ an optimal LCD code
and denote $d_{l}(n,k)=d_{o}(n,k)$. If $d_{l}(n,k)=d_{o}(n,k)-1$  ,
we call $\mathcal{C}$ an nearly optimal LCD code.

Defining the Hermitian inner product of ${\bf u}$, ${\bf v}\in$
$\mathbf{F}_{4}^{n}$ as
  $$({\bf u,v})={\bf uv^{\dag}}=u_{1}\bar{v_{1}}+u_{2}\bar
{v_{2}}+\cdot\cdot\cdot+u_{n}\bar{v_{n}}.$$
 The Hermitian dual code
of $\mathcal{C}$ $=[n,k]$ is $\mathcal{C}$$^{\perp h}$ $=\{x\mid
(x,y)_{h}=0,\forall y\in \mathcal{C}\}$, and $\mathcal{C}$$^{\perp
h}$ $=[n,n-k]$.  A generator matrix $H=H_{(n-k)\times n}$ of
$\mathcal{C}$$^{\perp h}$ is called a parity check matrix of
$\mathcal{C}$. If $\mathcal{C}$ $\subseteq$ $\mathcal{C}$$^{\perp
h}$, $\mathcal{C}$ is called a {\it weakly self-orthogonal} code. If
$\mathcal{C}$ is a  self-orthogonal code then each generator matrix
$G$ of $\mathcal{C}$ must satisfy rank($GG^{\dagger})=0$, where
$G^{\dagger}$ is the conjugate transpose of $G$.

If $\mathcal{C}$ $\cap$ $\mathcal{C}$$^{\perp h}=\{0\}$, then
$\mathcal{C}$ (or $\mathcal{C}$$^{\perp h}$) is called  a {\it
quaternary Hermitian LCD} code, and each generator matrix $G$ of
$\mathcal{C}$ must satisfy $k=$ rank($GG^{\dagger})$, see
Refs.\cite{Massey1992,Lu2014}.

In the following sections, we will discuss construction  of
Hermitian quaternary  LCD code $\mathcal{C}$ $=[n,k,d]$ with $d$ as
large as possible for given $n$ and $k\leq 5$. Firstly, we make some
notations for later use.

Let $\bf{1_{n}}$=$(1,1,...,1)_{1\times n}$  and
$\bf{0_{n}}$=$(0,0,...,0)_{1\times n}$ denote the all-one vector
 and  the all-zero vector of length $n$, respectively.
Construct
$$S_{2}=\left(
\begin{array}{ccccc}
0&1&1&1&1\\
1&0&1&\omega&\varpi\\
\end{array}
\right)=(\alpha_{1},...,\alpha_{5}),$$
 $$ S_{3}=
\left(
\begin{array}{cccccc}
S_{2}&\mathbf{0}_{2\times1}&S_{2}&S_{2}&S_{2}\\
\mathbf{0}_{5}&1&\mathbf{1}_{5}&\omega\mathbf{1}_{5}&\varpi\mathbf{1}_{5}\\
\end{array}
\right)=(\beta_{1},\beta_{2},\cdots,\beta_{21}),$$
 $$S_{4}=
\left(
\begin{array}{cccccc}
S_{3}&\mathbf{0}_{3\times1}&S_{3}&S_{3}&S_{3}\\
\mathbf{0}_{21}&1&\mathbf{1}_{21}&\omega\mathbf{1}_{21}&\varpi\mathbf{1}_{21}\\
\end{array}
\right)=(\gamma_{1},\gamma_{2},\cdots,\gamma_{85}).$$

 $$S_{5}=
\left(
\begin{array}{cccccc}
S_{4}&\mathbf{0}_{4\times1}&S_{4}&S_{4}&S_{4}\\
\mathbf{0}_{85}&1&\mathbf{1}_{85}&\omega\mathbf{1}_{85}&\varpi\mathbf{1}_{85}\\
\end{array}
\right)=(\zeta_{1},\zeta_{2},\cdots,\zeta_{341}).$$

$$\vdots$$

 $$S_{k}=
\left(
\begin{array}{cccccc}
S_{k-1}&\mathbf{0}_{k-1\times 1}&S_{k-1}&S_{k-1}&S_{k-1}\\
\mathbf{0}_{\frac{4^{(k-1)}}{3}}&1&\mathbf{1}_{\frac{4^{(k-1)}}{3}}&\omega\mathbf{1}_{\frac{4^{(k-1)}}{3}}&\varpi\mathbf{1}_{\frac{4^{(k-1)}}{3}}\\
\end{array}
\right)$$

It is well known that the matrix $S_{2}$ generates the $[5,2,4]$
simplex code with weight polynomial $1+15y^{4}$,  $S_{3}$ generates
the $[21,3,16]$ simplex code with weight polynomial $1+63y^{16}$,
$S_{4}$ generates  the $[85,4,64]$ simplex code with weight
polynomial $1+255y^{64}$, $S_{5}$ generates  the $[341,5,256]$
simplex code with weight polynomial $1+1023y^{256}$, and
$S_{k}S_{k}^{\dagger}=0$ for $k=2,3,4,5,\cdots$, see
Ref.\cite{MacWilliams,Li2004}.

\section{Hermitian LCD linear codes over $F_{4}$}
\noindent In this subsection, we construct optimal or near-optimal
Hermitian LCD codes over $F_{4}$. For $k\leq 5$, many optimal or
near-optimal Hermitian LCD codes are presented by Lu et al. in
Ref.\cite{Lu2014}.

\subsection{Constructing quaternary Hermitian LCD  codes  by  puncturing, shortening, extending and combination codes.
}

In this section, we discuss construction of $[n,k]$ Hermitian
Quaternary LCD codes for $k\geq 5$ and $n\geq 20$ from known codes
in [21, 23, 26]. The discussion is presented in four cases.

{\bf Lemma 1}\cite{Lu2014,Lu2017}: If $21\leq n\leq 25$ and $1
\leq k \leq 5$. Then $d_{l}(2,24)=18$, $d_{l}(2,25)=19$,
$d_{l}(3,21)=15$, $d_{l}(3,22)=15$, $d_{l}(4,22)=14$,
$d_{l}(4,23)=15$, $d_{l}(5,24)=15$. All the other Hermitian
Quaternary LCD codes are saturating the lower bound of Grassl's
codetable\cite{Grassl}.

{\bf proof:}  Ref.\cite{Lu2014,Lu2017} proved this lemma.

{\bf Theorem 2:} If $21\leq n\leq 25$ and $13\leq k\leq 18$, then
there exist 29 optimal quaternary Hermitian LCD codes saturating the
lower bound of Grassl's codetable\cite{Grassl} as in Table 1.

\begin{center}
 \noindent Table 1.  Optimal quaternary Hermitian LCD codes with $21\leq n\leq 25$ and $13\leq k \leq 19$.  \\
     [3mm]{ $ \begin{tabular}
{llllllllllllllllllllllll} \hline $\scriptstyle{n\setminus k}$

             &13    &14       &15        &16       &17       &18  &19     \\
\hline
21           &6     &5      &5         &4        &3         &2     &2     \\
22           &6     &6       &5       &4      &4         &3      &2        \\
23           &      &6      &6         &5      &4       &4        &3   \\
24           &      &        &6      &6        &5       &4       &4    \\
25           &      &         &       &6      &6         &5     &4     \\
\hline
\end{tabular}$
}
\end{center}

{\bf proof:} For $21 \leq n> 25$, calculating by Magma, one can
obtain 9 optimal  LCD codes are as follows: $[21, 14, 5]$, $[21, 15,
5]$, $[21, 16, 4]$, $[22, 16, 4]$, $[24, 16, 6]$, $[22, 18, 3]$,
$[23, 18, 4]$, $[25, 18, 5]$, $[23, 19, 3]$.

And then, calculating by Magma, one can obtain another 5 optimal
codes $[27, 19, 6]$, $[26, 21, 4]$, $[25, 18, 5]$, $[26, 20, 4]$,
$[33, 24, 6]$, which are not all quaternary Hermitian LCD codes.

Case 1. Construction of quaternary Hermitian LCD codes from
Puncturing:  Puncturing the $\mathcal{C}=$$[23, 15, 6]$on coordinate
sets $\{ 16 \}$, one can obtain $[22, 15, 5]$ Hermitian quaternary
LCD code.  Puncturing the $\mathcal{C}=$$[23, 17, 4]$on coordinate
sets  $\{ 1,19 \}$, one can obtain $[21, 17, 3]$ Hermitian
quaternary LCD code. Puncturing $\mathcal{C}=$$[24, 17, 5 ]$on
coordinate sets $\{1, 18 \}$, one can obtain $[22, 17, 4]$ Hermitian
quaternary  LCD code. Puncturing the $\mathcal{C}=$$[24, 19, 4 ]$on
coordinate sets $\{1, 4, 8 \}$, one can obtain $[21, 19, 2]$
quaternary Hermitian LCD code.

Case 2. Construction of LCD codes from Shortening: Shortening
$\mathcal{C}=$$[27, 19, 6]$ on coordinate sets
 $\{1,4 \}$, one can  obtain $\mathcal{C}=$$[25, 17, 6]$.
 Shortening  $\mathcal{C}=$$[27, 19, 6]$  on  coordinate sets
 $\{1,2,3,8 \}$, one can obtain $\mathcal{C}=$$[23, 15, 6]$.
 Shortening  $\mathcal{C}=$$[27, 19, 6]$  on  coordinate sets
 $\{1,2,3,4,7 \}$, one can obtain $\mathcal{C}=$$[22, 14, 6]$.
 Shortening  $\mathcal{D}=$$[27, 19, 6]$  on  coordinate sets
 $\{1,2,3,4,7,8 \}$, one can  obtain $\mathcal{C}=$$[21, 13, 6]$.
Shortening  $\mathcal{D}=$$[26, 21, 4]$ on  coordinate sets
 $\{1,2\}$  obtain $\mathcal{C}=$$[24, 19, 4]$. Shortening  $\mathcal{D}=$$[26, 21, 4]$ on  coordinate sets
 $\{1\}$  obtain $\mathcal{C}=$$[25, 20, 4]$.

 Shortening  $\mathcal{D}=$$[25, 18, 5]$  on  coordinate sets
 $\{4\}$, one can obtain $\mathcal{C}=$$[24, 17, 5]$.
 Shortening  $\mathcal{D}=$$[25, 18, 5]$  on  coordinate sets
 $\{1,4\}$, one can  obtain $\mathcal{C}=$$[23, 16, 5]$.

 Shortening  $\mathcal{D}=$$[26, 20, 4]$  on  coordinate sets
 $\{1\}$, one can  obtain $\mathcal{C}=$$[25, 19, 4]$.
 Shortening  $\mathcal{D}=$$[26, 20, 4]$  on  coordinate sets
 $\{1,2\}$, one can obtain $\mathcal{C}=$$[24, 18, 4]$.
 Shortening  $\mathcal{D}=$$[26, 20, 4]$  on  coordinate sets
 $\{1,2,4\}$, one can obtain $\mathcal{C}=$$[23, 17, 4]$.
  Shortening  $\mathcal{D}=$$[33, 24, 6]$  on  coordinate sets
 $\{1,2,3,4,5,6,7,8\}$, one can obtain $\mathcal{C}=$$[25, 16, 6]$.
 Shortening  $\mathcal{D}=$$[33, 24, 6]$  on  coordinate sets
 $\{1,2,3,4,5,6,7,8,9\}$, one can obtain $\mathcal{C}=$$[24, 15,
 6]$.
 Shortening  $\mathcal{D}=$$[33, 24, 6]$  on  coordinate sets
 $\{1,2,3,4,5,6,7,8,9,10\}$, one can obtain $\mathcal{C}=$$[23, 14,
 6]$.
 Shortening  $\mathcal{D}=$$[33, 24, 6]$   on  coordinate sets
 $\{1,2,3,4,5,6,7,8,9,10,14\}$, one can obtain $\mathcal{C}=$$[22, 13,
 6]$.

{\bf  Remark 3.1}  In Theorem 2, all of the codes are optimal
quaternary Hermitian LCD codes. Since $[21,3,16]$ is a Simplex code,
there is no $[21,18,3]$ quaternary Hermitian LCD codes. Hence,
$[21,18,2]$ is optimal quaternary Hermitian LCD codes. For
shortening $\mathcal{D}=$$[33, 24, 6]$ on coordinate sets
 $\{1,2,3,4,5,6,7,8,9,10,11,12\}$, we can obtain $\mathcal{C}=$$[21, 12,
 6]$. It is nearly-optimal quaternary Hermitian LCD code with  weight enumerator
 $1+279z^{6}+1116z^{7}+5739z^{8}+22023z^{9}+79815z^{10}+...$.

{\bf Theorem 3:} If $21\leq n\leq 25$ and $6\leq k\leq 8$, then
$d_{l}(6,21)=12$, $d_{l}(6,22)=12$, $d_{l}(6,23)=13$,
$d_{l}(6,24)=14$, $d_{l}(8,25)=12$, $d_{l}(7,20)=10$. All this codes
are quaternary Hermitian LCD codes saturating the lower or upper
bound of Grassl's codetable.

{\bf proof:} A constacyclic code $\mathcal{C}$ $=[21,15,5]$ is given
in [21], where its generator polynomial is
$x^{6}+\overline{\omega}x^{5}+x^{4}+\overline{\omega}x^{2}+x+\overline{\omega}$
. The dual code of  $\mathcal{C}$  is the code $\mathcal{D}$ $=[21,
6, 12]$ with a generator matrix $G_{6,21}$, and both of
$\mathcal{C}$ and $\mathcal{D}$ are quaternary Hermitian LCD codes.

Let

$G_{6,21}=\left(
\begin{array}{ccccc}
2 1 1 2 1 0 2 2 1 1 0 2 1 2 2 1 0 0 0 0 0 \\
3 0 3 2 0 1 3 1 0 3 1 3 0 2 1 0 1 0 0 0 0 \\
2 2 1 1 3 0 3 1 0 1 3 3 2 2 0 0 0 1 0 0 0 \\
0 2 2 1 1 3 0 3 1 0 1 3 3 2 2 0 0 0 1 0 0 \\
3 2 0 1 3 1 0 3 1 3 0 2 1 0 1 0 0 0 0 1 0 \\
2 2 3 2 0 3 3 2 2 0 3 2 3 3 2 0 0 0 0 0 1 \\
\end{array}\right)$,  $G_{6,24}=\left(\begin{array}{cccccc}
0 2 2 3 1 2 1 1 0 0 0 3 2 1 1 1 0 0 0 0 0 3 2 1 \\
1 0 2 2 3 1 2 1 1 0 0 0 3 2 1 1 1 0 0 0 0 0 3 2 \\
1 1 0 2 2 3 1 2 1 1 0 0 0 3 2 1 2 1 0 0 0 0 0 3 \\
2 1 1 0 2 2 3 1 1 1 1 0 0 0 3 2 3 2 1 0 0 0 0 0 \\
1 2 1 1 0 2 2 3 2 1 1 1 0 0 0 3 0 3 2 1 0 0 0 0 \\
3 1 2 1 1 0 2 2 3 2 1 1 1 0 0 0 0 0 3 2 1 0 0 0 \\
2 3 1 2 1 1 0 2 0 3 2 1 1 1 0 0 0 0 0 3 2 1 0 0
\\\end{array}\right)$,

$G_{7,20}=\left(\begin{array}{cccccc}
1 1 0 1 1 1 1 1 1 0 1 0 0 0 0 0 0 0 1 1 \\
1 1 1 2 0 3 2 3 0 0 3 1 1 0 0 0 0 0 1 2 \\
3 0 1 3 2 2 2 3 0 1 1 0 3 0 0 0 0 0 2 0 \\
1 1 2 0 2 0 2 3 0 0 1 0 0 1 1 0 0 0 0 1 \\
1 1 0 0 0 1 3 2 0 0 2 0 1 3 0 1 0 0 0 1 \\
3 0 2 2 0 2 3 3 0 0 0 0 1 3 0 0 1 0 0 3 \\
3 0 3 1 3 0 0 2 0 0 3 0 1 2 0 0 0 1 0 1 \\\end{array}\right)$,
$G_{8,25}=\left(\begin{array}{cccccc}
1 0 0 0 0 0 0 0 1 0 0 1 2 1 2 3 2 1 3 1 0 3 3 1 0\\
0 1 0 0 0 0 0 0 3 1 3 2 2 1 1 1 1 2 3 2 0 3 3 1 1\\
0 0 1 0 0 0 0 0 1 2 1 2 0 3 3 2 3 0 3 2 2 3 0 2 1\\
0 0 0 1 0 0 0 0 2 1 1 2 2 0 0 1 1 0 0 0 2 2 3 0 3\\
0 0 0 0 1 0 0 0 1 3 3 3 2 2 2 1 3 3 0 2 0 2 2 3 3\\
0 0 0 0 0 1 0 0 1 0 1 1 3 2 0 3 3 1 0 2 2 0 2 2 0\\
0 0 0 0 0 0 1 0 0 1 0 1 1 3 2 0 3 3 0 0 2 2 0 2 2\\
0 0 0 0 0 0 0 1 1 2 2 0 0 2 1 2 2 3 2 0 0 0 0 3 3\\
\end{array}\right)$,

 There exists a quaternary Hermitian LCD code $[24,6,14]$ with generator matrix
 $G_{6,24}$. Its weight enumerator  is
$1+207z^{14}+378z^{15}+630z^{16}+360z^{17}+495z^{18}+1062z^{19}+585z^{20}+180z^{21}+162z^{22}+36z^{23}$.
Puncturing $\mathcal{C}$$=[24,6,14]$ on coordinate sets $\{7\}$,
$\{1,3\}$, , we can obtain two quaternary Hermitian LCD codes
$[23,6,13]$, $[22,6,12]$.

 There exists a quaternary Hermitian LCD code $[20,7,10]$ with generator
matrix
 $G_{7,20}$. Its weight enumerator  is
$1+210z^{10}+594z^{11}+969z^{12}+1647z^{13}+2703z^{14}+3519z^{15}+3060z^{16}+2205z^{17}+1107z^{18}+291z^{19}+78z^{20}$.

 There
exists a quaternary Hermitian LCD code $[25, 8, 12]$ with generator
matrix
 $G_{8,25}$. Its weight enumerator  is
$1+177z^{12}+540z^{13}+1365z^{14}+2721z^{15}+4836z^{16}
 +8283z^{17}+10938z^{18}
 +11694z^{19}+10983z^{20}+7734z^{21}+4185z^{22}+1617z^{23}+411z^{24}+25z^{25}$.

 Shortening the $[25, 8, 12]$ quaternary Hermitian LCD code   on  coordinate sets
 $\{2\}$, one can obtain $[24, 7,
 12]$. Its weight enumerator  is
$1+102z^{12}+267z^{13}+561z^{14}+1086z^{15}+1764z^{16}
 +2628z^{17}+3144z^{18}
 +2730z^{19}+2226z^{20}+1233z^{21}+495z^{22}+120z^{23}+27z^{24}$.
We can deduce a submatrices $G_{7,25}$ form $G_{8,25}$. $G_{7,25}$
as a generator matrix, one can obtain $[25, 7,
 12]$.

{\bf Theorem 4:}  If $21\leq n\leq 25$ and $8\leq k\leq 15$, then
there exist 27 quaternary Hermitian LCD codes as in Table 2.

\begin{center}
 \noindent Table 2. quaternary Hermitian LCD codes with $21\leq n\leq 25$ and $8\leq k \leq 15$.  \\
     [3mm]{ $ \begin{tabular}
{llllllllllllllllllllllll} \hline $\scriptstyle{n\setminus k}$

             &8    &9       &10        &11       &12        &13     &14     &15  \\
\hline
21           &8      &8       & 7         &6         &          &     &       &   \\
22           &9      &8       &8          &7         &6          &     &       &      \\
23           &10      &9       &8          &8        &7          & 6     &       &  \\
24           &11      &10       &9          &8         &8          &7      &6       &    \\
25           &      &11       &10          &9         &8          & 7      &7      &6   \\
\hline
\end{tabular}$
}
\end{center}

 {\bf proof:}  Let

 $A_{12}^{\top}=\left(\begin{array}{cccccc}
3 0 0 2 0 1 2 0 0 1 2 0 3 0 0 3 1 2 \\
3 3 0 3 2 2 2 2 0 2 2 2 1 3 0 1 1 2 \\
1 3 3 3 3 0 1 2 2 2 1 2 3 1 3 1 3 2 \\
2 1 3 0 3 1 3 1 2 0 1 1 3 3 1 2 3 0 \\
0 0 2 0 3 3 2 1 3 2 3 0 3 1 3 1 2 3 \\
0 2 1 3 0 3 1 3 1 2 0 1 1 3 3 1 2 3 \\
3 0 0 3 0 0 2 2 1 0 3 3 2 3 1 1 2 3 \\
3 3 0 1 3 3 1 2 2 2 1 3 1 2 3 3 2 3 \\
0 3 3 1 1 0 2 1 2 1 3 1 1 1 2 1 0 3 \\
2 0 3 2 1 2 1 2 1 1 0 3 3 1 1 0 2 1 \\
0 2 0 1 2 0 0 1 2 0 3 0 0 3 1 2 1 0 \\
0 0 2 0 1 2 0 0 1 2 0 3 0 0 3 1 2 1 \\\end{array}\right)$,
$A_{11}^{\top}=\left(\begin{array}{cccccc}
1 2 3 2 1 3 0 0 3 0 3 3 2 1 0 0 \\
3 2 0 2 3 3 1 1 3 1 1 0 2 1 3 2 \\
1 1 0 0 1 3 2 3 0 2 3 2 0 2 0 0 \\
1 3 2 1 3 3 2 2 3 1 2 2 3 3 2 0 \\
0 0 1 3 2 1 3 2 0 3 3 0 3 0 0 2 \\
0 1 3 1 0 2 1 1 1 0 3 3 0 3 2 3 \\
1 3 2 2 3 2 2 2 2 1 2 1 2 3 3 3 \\
2 2 0 3 0 1 2 1 1 2 3 0 0 1 3 2 \\
3 1 2 2 1 3 1 0 1 1 1 0 2 1 2 0 \\
0 0 0 2 2 1 2 1 2 0 3 2 0 2 2 2 \\
3 2 2 3 3 3 1 0 3 2 1 2 1 2 2 1 \\\end{array}\right)$.

Let $A_{10}^{\top}=\left(\begin{array}{cccccc}
1 1 1 2 1 1 2 2 1 2 0 2 3 0 3 2 3 2 1 0 \\
0 0 0 0 0 0 0 0 0 0 0 0 0 0 0 0 0 0 0 0 \\
1 2 1 0 3 3 1 3 2 0 1 0 0 3 1 0 2 1 1 1 \\
1 1 3 3 1 3 1 3 0 3 2 0 0 0 1 1 2 0 3 2 \\
2 1 1 3 3 3 2 0 0 0 3 3 0 0 0 0 3 2 2 0 \\
0 1 0 3 0 3 2 2 2 0 3 2 3 3 2 2 2 1 2 1 \\
3 2 2 1 0 1 2 0 1 1 3 3 0 0 1 1 2 0 3 2 \\
0 0 3 3 1 2 1 2 2 0 1 2 1 1 2 3 0 3 2 0 \\
1 1 3 2 1 3 0 2 0 1 3 2 3 2 2 1 0 0 0 3 \\
1 1 0 1 0 1 1 0 2 1 2 0 2 3 0 2 0 3 2 3 \\\end{array}\right)$,
$B_{16}=\left(\begin{array}{cccccc}
 0 3 3 1 2 0 0 3 3 0 2 0 0\\
 0 0 3 3 1 0 2 0 3 3 0 2 0\\
 2 3 3 0 3 3 0 1 1 2 1 0 0\\
 0 2 3 3 0 0 3 3 1 1 2 1 0\\
 1 2 0 1 3 0 3 3 0 2 0 2 0\\
 2 2 1 3 1 2 2 1 2 1 0 0 0\\
 0 2 2 1 3 2 1 2 1 2 1 0 0\\
 0 0 2 2 1 1 3 2 2 1 2 1 0\\
 1 2 2 0 2 0 2 2 1 1 0 2 0\\
 2 2 1 1 0 3 3 1 3 0 3 0 0\\
 0 2 2 1 1 3 0 3 1 3 0 3 0\\
 3 1 3 3 1 2 3 1 1 3 0 0 0\\
 0 3 1 3 3 3 1 2 1 1 3 0 0\\
 0 0 3 1 3 1 3 3 2 1 1 3 0\\
 3 1 1 2 1 1 1 3 1 0 2 1 0\\
 1 1 3 3 2 2 2 2 0 2 1 2 0\\\end{array}\right)$,

There exists a
code $[30, 18, 8]$ with generator matrix
 $G_{18,30}=\bigg[I_{18} \bigg | A_{12}\bigg]$. It is not a quaternary Hermitian LCD code.
 Shortening  $\mathcal{C}=$$[30, 18, 8]$  on  coordinate sets
 $\{ 1, 3, 6, 12, 13, 17 \}$, one can obtain a quaternary Hermitian LCD code $[24, 12, 8]$.
  Shortening  $\mathcal{C}=$$[30, 18, 8]$  on  coordinate sets
 $\{1,2,3,4,5,6,8\}$, one can obtain a quaternary Hermitian LCD code $[23, 11, 8]$.
 Shortening  $\mathcal{C}=$$[30, 18, 8]$  on  coordinate sets
 $\{1,2,3,4,5,6,7,8\}$, one can obtain a quaternary Hermitian LCD code $[22, 10, 8]$.
 Shortening  $\mathcal{C}=$$[30, 18, 8]$  on  coordinate sets
 $\{1,2,3,4,5,6,7,8,10\}$, one can obtain a quaternary Hermitian LCD code $[21, 9, 8]$.

 There exists
a code $[27, 16, 7]$ with generator matrix
 $G_{16,27}=\bigg[I_{16} \bigg | A_{11}\bigg]$.
 Shortening  $\mathcal{C}=$$[27, 16, 7]$  on  coordinate sets
 $\{1,2\}$, one can obtain a quaternary Hermitian LCD code $[25, 14, 7]$.
 Shortening  $\mathcal{C}=$$[27, 16, 7]$  on  coordinate sets
 $\{1,2,3\}$, one can obtain a quaternary Hermitian LCD code $[24, 13, 7]$.
  Shortening  $\mathcal{C}=$$[27, 16, 7]$  on  coordinate sets
 $\{1,2,3,4\}$, one can obtain a quaternary Hermitian LCD code $[23, 12, 7]$.
 $\mathcal{C}=$$[27, 16, 7]$ on  coordinate sets
 $\{1,2,3,4,5\}$, one can obtain a quaternary Hermitian LCD code $[22, 11, 7]$.
  Shortening  $\mathcal{C}=$$[27, 16, 7]$  on  coordinate sets
 $\{1,2,3,4,5,6\}$, one can obtain a quaternary Hermitian LCD code $[21, 10, 7]$.

There exists a code $[30, 20, 6]$ with generator matrix
 $G_{20,30}=\bigg[I_{20} \bigg | A_{10}\bigg]$.
  Shortening  $\mathcal{C}=$$[30, 20, 6]$  on  coordinate sets
 $\{1,2,3,4,5\}$, one can obtain a  quaternary Hermitian  LCD code $[25, 15, 6]$.
  Shortening  $\mathcal{C}=$$[30, 20, 6]$ stored in Magma on  coordinate sets
 $\{1,2,3,4,5,6\}$, one can obtain a quaternary Hermitian LCD code $[24, 14, 6]$.
   Shortening  $\mathcal{C}=$$[30, 20, 6]$  on  coordinate sets
 $\{1,2,3,5,6,7,13\}$, one can obtain a quaternary Hermitian LCD code $[23, 13, 6]$.
    Shortening  $\mathcal{C}=$$[30, 20, 6]$ stored in Magma on  coordinate sets
 $\{1,2,3,4,5,6,7,8\}$, one can obtain a  quaternary Hermitian LCD code $[22, 12, 6]$.
    Shortening  $\mathcal{C}=$$[30, 20, 6]$ stored in Magma on  coordinate sets
 $\{1,2,3,4,5,6,7,8,9\}$, one can obtain a quaternary Hermitian LCD code $[21, 11, 6]$.

There exists a code $[29, 16, 8]$ with generator matrix
 $G_{16,29}=\bigg[I_{16} \bigg | B_{16}\bigg]$. It is not a quaternary Hermitian  LCD
 code.
Shortening  $\mathcal{C}=$$[29, 16, 8]$  on coordinate sets
 $\{1,2,3,4\}$, one can obtain a quaternary Hermitian LCD code $[25, 12, 8]$.
    Shortening  $\mathcal{C}=$$[29, 16, 8]$ stored in Magma on  coordinate sets
 $\{1,2,3,4,6\}$, one can obtain a quaternary Hermitian LCD code $[24, 11, 8]$.
    Shortening  $\mathcal{C}=$$[29, 16, 8]$ stored in Magma on  coordinate sets
 $\{1,2,3,4,5,6\}$, one can obtain a quaternary Hermitian  LCD code $[23, 10, 8]$.
    Shortening  $\mathcal{C}=$$[29, 16, 8]$ stored in Magma on  coordinate sets
 $\{1,2,3,4,5,6,8\}$, one can obtain a quaternary Hermitian  LCD code $[22, 9, 8]$.
  Shortening  $\mathcal{C}=$$[29, 16, 8]$ stored in Magma on  coordinate sets
 $\{1,2,3,4,5,6,7,8\}$, one can obtain a quaternary Hermitian  LCD code $[21, 8, 8]$.

Let

$A_{14}=\left(\begin{array}{cccccc}
1 3 0 0 0 1 2 3 0 0 2 2 1 1 \\
1 0 2 1 1 1 0 3 2 1 1 3 3 0 \\
0 1 0 2 1 1 1 0 3 2 1 1 3 3 \\
3 3 2 3 1 2 2 2 3 0 1 2 2 0 \\
0 3 3 2 3 1 2 2 2 3 0 1 2 2 \\
2 2 1 1 0 1 3 0 0 0 1 2 3 0 \\
0 2 2 1 1 0 1 3 0 0 0 1 2 3 \\
3 3 1 1 2 2 3 2 0 3 3 3 2 1 \\
1 2 2 0 0 3 3 2 3 1 2 2 2 3 \\
3 2 1 1 3 3 0 0 1 0 2 1 1 1 \\
1 2 3 0 0 2 2 1 1 0 1 3 0 0 \\
0 1 2 3 0 0 2 2 1 1 0 1 3 0 \\
0 0 1 2 3 0 0 2 2 1 1 0 1 3 \\
3 3 3 2 1 0 3 3 1 1 2 2 3 2 \\
2 1 1 1 0 3 2 1 1 3 3 0 0 1 \\
1 1 1 1 1 1 1 1 1 1 1 1 1 1 \\\end{array}\right)$,
$A_{15}=\left(\begin{array}{cccccc}
3 0 1 3 2 2 2 0 2 1 1 3 1 1 3 \\
2 1 1 2 2 1 1 3 1 1 1 2 3 3 1 \\
3 1 3 0 0 3 0 3 1 0 1 0 2 0 3 \\
2 1 0 0 1 3 0 1 2 2 0 2 0 0 0 \\
0 2 1 0 0 1 3 0 1 2 2 0 2 0 0 \\
0 0 2 1 0 0 1 3 0 1 2 2 0 2 0 \\
0 0 0 2 1 0 0 1 3 0 1 2 2 0 2 \\
1 1 3 2 1 3 2 3 2 1 0 3 2 3 0 \\
0 1 1 3 2 1 3 2 3 2 1 0 3 2 3 \\
2 2 0 2 2 1 2 2 3 0 2 2 0 1 2 \\
1 3 1 2 1 0 3 1 1 1 0 0 2 1 1 \\
3 2 1 0 0 0 1 1 3 0 1 1 0 1 1 \\
3 0 0 0 2 1 1 3 3 2 0 0 1 3 1 \\
3 0 2 1 2 3 0 3 1 2 2 1 0 2 3 \\
2 1 1 1 0 1 0 1 2 2 2 1 1 2 2 \\
1 3 2 3 2 2 3 3 2 0 2 0 1 0 2 \\\end{array}\right)$,
$A_{12}=\left(\begin{array}{cccccc}
3 3 3 2 3 1 3 0 2 0 1 2 \\
1 2 2 0 2 1 2 3 0 2 3 3 \\
0 1 2 2 3 1 3 3 1 3 3 3 \\
1 1 0 1 2 1 3 3 3 1 0 1 \\
3 2 2 2 1 3 1 2 3 3 3 1 \\
1 2 3 1 3 3 3 2 3 3 2 3 \\
3 2 1 1 3 1 1 1 1 0 2 1 \\
3 0 1 3 3 1 3 2 2 2 1 1 \\
2 1 2 0 3 3 0 2 3 1 3 3 \\
0 2 1 2 3 2 0 1 3 2 0 0 \\
1 1 3 2 0 2 2 1 2 0 2 0 \\
0 0 0 0 1 2 1 0 2 2 2 1 \\
0 0 0 0 0 1 1 1 0 2 2 2 \\\end{array}\right)$,

There exists a code $[30, 16, 9]$ with generator matrix
 $G_{16,30}=\bigg[I_{16} \bigg | A_{14}\bigg]$. It is not a quaternary Hermitian  LCD
 code.    Shortening  $\mathcal{C}=$$[30, 16, 9]$  on  coordinate sets
 $\{1,2,3,4,5\}$, one can obtain  a quaternary Hermitian  LCD code $[25, 11, 9]$.
     Shortening  $\mathcal{C}=$$[30, 16, 9]$ a on  coordinate sets
 $\{1,2,3,4,5,11\}$, one can obtain a quaternary Hermitian   LCD code $[24, 10, 9]$.
      Shortening  $\mathcal{C}=$$[30, 16, 9]$  on  coordinate sets
 $\{1,2,3,4,5,6,11\}$, one can obtain a quaternary Hermitian  LCD code $[23, 9, 9]$.
     Shortening  $\mathcal{C}=$$[30, 16, 9]$  on  coordinate sets
 $\{1,2,3,4,5,6,11\}$, one can obtain a quaternary Hermitian  LCD code $[22, 8, 9]$.

 There exists a
code $[31, 16, 10]$ with generator matrix
 $G_{16,31}=\bigg[I_{16} \bigg | A_{15}\bigg]$. It is not a quaternary Hermitian  LCD
 code.        Shortening $\mathcal{C}=$$[31, 16, 10]$  on  coordinate sets
 $\{1,2,3,4,11\}$, one can obtain a quaternary Hermitian  LCD
 code$[26, 11, 10]$.     Shortening $\mathcal{C}=$$[31, 16, 10]$  on  coordinate sets
 $\{1,2,3,4,11,17\}$, one can obtain a quaternary Hermitian  LCD code$[25, 10, 10]$.
        Shortening $\mathcal{C}=$$[31, 16, 10]$ on  coordinate sets
 $\{1,2,3,4,8,11,17\}$, one can obtain a quaternary Hermitian  LCD code$[24, 10, 10]$.
        Shortening $\mathcal{C}=$$[31, 16, 10]$  on  coordinate sets
 $\{1,2,3,4,8,11,17,18\}$, one can obtain a quaternary Hermitian  LCD code$[23, 10, 10]$.

There exists a quaternary Hermitian  LCD code $[25, 13, 7]$ with
generator matrix
 $G_{13,25}=\bigg[I_{13} \bigg | A_{12}\bigg]$.

{\bf Theorem 5:} $d_{l}(7,21)=10$,
$d_{l}(7,22)=11$,$d_{l}(7,22)=11$, $d_{l}(7,23)=12$,
$d_{l}(7,25)=13$, $d_{l}(6,25)=14$, $d_{l}(12,25)=8$,
$d_{l}(13,25)=7$, $d_{l}(7,18)=9$ and $d_{l}(7,19)=9$ are Hermitian
quaternary  LCD codes.

 {\bf proof:}  Let

 $G_{7,24}=\left(\begin{array}{cccccc}
0 2 2 3 1 2 1 1 0 0 0 3 2 1 1 1 0 0 0 0 0 3 2 1 \\
1 0 2 2 3 1 2 1 1 0 0 0 3 2 1 1 1 0 0 0 0 0 3 2 \\
1 1 0 2 2 3 1 2 1 1 0 0 0 3 2 1 2 1 0 0 0 0 0 3 \\
2 1 1 0 2 2 3 1 1 1 1 0 0 0 3 2 3 2 1 0 0 0 0 0 \\
1 2 1 1 0 2 2 3 2 1 1 1 0 0 0 3 0 3 2 1 0 0 0 0 \\
3 1 2 1 1 0 2 2 3 2 1 1 1 0 0 0 0 0 3 2 1 0 0 0 \\
2 3 1 2 1 1 0 2 0 3 2 1 1 1 0 0 0 0 0 3 2 1 0 0
\\\end{array}\right)$,
  $G_{7,25}=\left(\begin{array}{cccccc}
1 1 0 0 0 0 0 0 3 2 0 3 1 3 0 0 1 3 1 3 0 3 1 2 1 \\
0 0 1 0 0 0 0 0 0 3 2 0 3 1 3 0 2 1 3 1 3 0 3 1 1 \\
0 0 0 1 0 0 0 0 0 3 1 2 3 2 2 3 3 3 2 2 2 3 3 2 1 \\
0 0 0 0 1 0 0 0 3 3 1 1 1 2 1 2 0 2 0 3 1 2 0 2 1 \\
0 0 0 0 0 1 0 0 2 0 1 1 2 0 1 1 0 1 1 1 0 1 1 1 1 \\
0 0 0 0 0 0 1 0 1 1 2 1 2 3 3 1 3 1 2 0 2 0 2 0 1 \\
0 0 0 0 0 0 0 1 1 2 3 2 2 3 0 3 2 2 2 3 3 2 3 3 1
\\\end{array}\right)$,

$G_{7,19}=\left(\begin{array}{cccccc}
1 1 3 0 1 2 1 2 3 1 0 1 0 0 0 0 0 0 0 \\
0 1 3 3 2 0 3 3 2 0 0 3 1 2 0 0 0 0 0 \\
0 3 0 3 3 3 2 3 2 0 3 1 0 1 0 0 0 0 0 \\
0 1 3 1 0 3 0 3 2 0 0 1 0 0 1 1 0 0 0 \\
0 1 3 0 0 0 1 1 1 0 0 2 0 2 3 0 3 0 0 \\
0 3 0 1 2 0 2 1 2 0 0 0 0 2 3 0 0 1 0 \\
0 3 0 2 1 1 0 0 1 0 0 3 0 2 2 0 0 0 3 \\\end{array}\right)$,
$G_{6,25}=\left(\begin{array}{cccccc}
3 1 1 1 2 0 1 3 3 0 1 0 0 0 0 1 2 3 0 1 3 3 1 2 2 \\
1 1 1 1 1 1 0 1 3 3 0 3 0 0 0 0 1 2 2 0 1 3 3 1 2 \\
2 3 1 1 1 0 0 0 1 3 3 2 3 0 0 0 0 1 2 2 0 1 3 3 1 \\
1 1 3 1 1 3 1 0 0 1 3 1 2 3 0 0 0 0 1 2 2 0 1 3 3 \\
1 2 1 3 1 3 0 1 0 0 1 0 1 2 3 0 0 0 3 1 2 2 0 1 3 \\
1 1 2 1 3 1 3 0 1 0 0 0 0 1 2 3 0 0 3 3 1 2 2 0 1
\\\end{array}\right)$

 There exists a
code $[31, 16, 10]$ with generator matrix
 $G_{7,24}$. Its weight enumerators is
$1+384z^{13}+744z^{14}+888z^{15}+1746z^{16}+2544z^{17}+3156z^{18}+2928z^{19}+2118z^{20}+1200z^{21}+540z^{22}+120z^{23}+15z^{24}$.
It is not a quaternary Hermitian  LCD  code. Puncturing
$\mathcal{C}$$_{1}$ on coordinate sets $\{1\}$, $\{1,3\}$,
$\{1,2,7\}$, we can obtain $[23,7,12]$, $[22,7,11]$, $[21,7,10]$
quaternary Hermitian  LCD codes.

 There exists a quaternary Hermitian  LCD
code $[25, 7, 13]$ with generator matrix
 $G_{7,25}$. Its weight enumerators is
$1+189z^{13}+495z^{14}+750z^{15}+1179z^{16}+1908z^{17}+2577z^{18}+2967z^{19}+2667z^{20}+1932z^{21}+1092z^{22}+495z^{23}+117z^{24}+15z^{25}$.

There exists a quaternary Hermitian  LCD code $[25, 6, 14]$ with
generator matrix
 $G_{6,25}$. Its weight enumerators is
$1+48z^{14}+240z^{15}+432z^{16}+534z^{17}+573z^{18}+648z^{19}+657z^{20}+510z^{21}+363z^{22}+84z^{23}+6z^{24}$,

There exists  optimal  quaternary Hermitian  LCD codes  $[19, 7, 9]$
 with generator matrix $G_{7,19}$. Its weight enumerator is
 $1+195z^{9}+483z^{10}+888z^{11}+1479z^{12}+2361z^{13}+3165z^{14}+3327z^{15}+2508z^{16}+1368z^{17}+492z^{18}+117z^{19}$.
 Puncturing the quaternary Hermitian  LCD codes $[19, 7, 9]$   on coordinate sets
$\{1\}$, one can obtain optimal quaternary Hermitian  LCD codes
$[18, 7, 9]$ with weight enumerator
$1+393z^{9}+666z^{10}+1245z^{11}+2193z^{12}+3315z^{13}+3597z^{14}+2799z^{15}+1554z^{16}+504z^{17}+117z^{18}$,

\section{Conclusion}

In this paper, we studied  constructions of  quaternary Hermitian
LCD codes. For each $k\leq n$ and $n\leq 25$, we constructed an
$[n,k]$ quaternary Hermitian  LCD code. When constructing quaternary
Hermitian  LCD code by a given code, we tried all possible
coordinates and chose one case that result in quaternary Hermitian
LCD code with highest minimum distance as output. Some of these
quaternary Hermitian LCD code constructed in this paper are optimal
codes and saturate the bound on  the minimum distance of codetable
in \cite{Grassl}, and some of them  are nearly optimal codes.
According to weight enumerators for classification codes in
\cite{Bouyukliev}, there exist some optimal codes $[15,4,9]$,
$[16,4,10]$, $[17,4,11]$, $[19,4,14]$ , $[23,5,15]$ which are not
LCD codes. And the number of these code is one. Thus, the
$[15,4,8]$, $[16,4,9]$, $[17,4,10]$, $[19,4,13]$, $[23,5,14]$
quaternary Hermitian  LCD codes are optimal. In \cite{Bouyukliev},
there only exist three codes of  $[15,6,8]$ which all of them are
self-orthogonal. Thus, the quaternary Hermitian  LCD code
$[15,6,7]$ is optimal. We emphasize that there are three quaternary
Hermitian LCD $[18,7,9]$, $[19,7,9]$ and $[20,7,10]$ which are
 optimal.

According to \cite{Grassl}, the following quaternary Hermitian  LCD
codes we give in this section are also optimal linear codes: $[n,k,
d]$ for $21\leq n\leq 25$ and $16\leq k\leq 18$; $[n-i,15-i, d]$ for
$21\leq n\leq 24$ and $0\leq i\leq 2$;  $[23, 4, 15]$, $[24, 5,
15]$, $[21, 6, 12]$, $[22, 6, 12]$, $[23, 6, 13]$, $[24, 6, 14]$,
$[25, 8, 12]$ and $[20, 7, 10]$. Except these mentioned above codes,
the quaternary Hermitian  LCD codes constructed in this paper do not
attain known upper or lower bounds on the minimum distance of a
linear code. Nonetheless, the minimum distances of those codes
appear very good in general. These codes are the best possible among
those obtainable by our approach.

In Table 3,  many lower and upper bounds on minimal distance of
Hermitian LCD codes with length $n\geq 25$ are from known
constructed codes. To make the bounds in Table 3 tighter, we need to
choose  other quaternary Hermitian  LCD codes better than that given
in  this paper and investigate other code constructions to raise the
lower bounds.  We also plan to explore the construction of Hermitian
LCD codes from geometric view to decrease the upper bound.

 In \cite{Lu2014,lai2,lai3}, it has shown that
  if there exist a quaternary Hermitian  $[n,k,d$] code over $F_{q^{2}}$, then there
 exist a maximal entanglement Entanglement-assisted quantum error correcting code
(EAQECC) over $F_{q}$ with parameters $[[n, 2k- n + c, d; c]]$,
where c is the rank of the product of the parity check matrix and
its conjugate. Moreover, a maximal entanglement EAQECC derived from
a LCD code has the same minimum distance with the underlying
classical code. Hence, all of the optimal quaternary Hermitian LCD
codes can be used to constructed to optimal binary maximal
entanglement EAQECC. In addition, From three quaternary Hermitian
LCD codes $[18,7,9]$, $[19,7,9]$ and $[20,7,10]$ given in this
paper, maximal entanglement EAQECC $[[18,7,9;11]]$, $[[19,7,9;12]]$
and $[[20,7,10;12]]$ can be constructed, which improve the minimal
distance of the codes in\cite{Lu2014,lai3}.

\begin{center}
 \noindent Table 3. Lower and upper bounds on the minimum distance of
    Quaternary Hermitian LCD codes.  The bold face entries represent improvements over the prior works.\\
     [3mm]{ $ \begin{tabular}
{llllllllllllllllllllllll} \hline $\scriptstyle{n\setminus k}$

    &1 &2  &3     &4        &5        &6        &7         &8      &9      &10    &11     &12        \\
3  &3  &2  &      &         &         &          &        &        &       &      &       &      \\
4  &3  &2  &1     &         &         &          &        &        &       &      &       &     \\
5  &5  &3  &2     &2        &         &          &        &        &       &      &       &     \\
6  &5  &4  &3     &2        &1        &          &        &        &       &      &       &     \\
7  &7  &5  &4     &3        &2        &2         &        &        &       &      &       &     \\
8  &7  &6  &5     &4        &3        &2         &1       &        &       &      &       &     \\
9  &9  &6  &6     &5        &4        &3         &2       &2       &       &      &       &     \\
10 &9  &7  &6     &6        &5        &4         &3       &2       &1      &1     &       &    \\
11 &11 &8  &7     &6        &6        &5         &4       &3       &2      &2     &1      &     \\
12 &11 &9  &8     &7        &6        &5-6       &5       &4       &3      &2     &2      &1    \\
13 &13 &10 &9     &8        &7        &6         &5       &4       &4      &3     &3      &2     \\
14 &13 &10 &9     &8        &7-8      &6-7       &6       &5       &4      &4     &4      &2     \\
15 &15 &11 &10    &9       &8         &7       &6-7     &6       &5      &4     &4      &3     \\
16 &15 &12 &11    &10       &9        &8         &7-8     &6-7     &6      &5     &5      &4      \\
17 &17 &13 &12    &11       &9-10     &8-9       &7-8     &7-8     &6-7    &6     &5-6    &4            \\
18 &17 &14 &13    &11-12    &10-11    &9-10      &\bf{9}  &8-9     &7-8    &6-7   &5-6    &5        \\
19 &19 &14 &13    &12-13    &11       &10-11     &\bf{9}  &8-9     &8      &7     &6-7    &5-6      \\
20 &19 &15 &14    &13       &12       &11-12     &\bf{10} &8-9    &8-9    &7-8   &6-7    &6-7         \\
21 &21 &16 &15    &14       &12       &12        &10-11   &8-10    &8-9   &7-9    &6-8    &6-7       \\
22 &22 &17 &15    &14       &13       &12-13     &11-12   &9-10    &8-10   &8-9   &7-9    &6-8            \\
23 &23 &18 &16    &15       &14       &13        &12-13   &10-12   &9-11   &8-10  &8-9    &7-9       \\
24 &24 &18 &17    &16       &15       &14        &12-13   &11-13   & 10-12 &9-11   &8-10  &8-9    \\
25 &25 &19 &18    &17       &15       &14-15    &13-14   &12-13    & 11-13  &10-12 &9-11  &8-10           \\
\hline
\end{tabular}$
}
\end{center}

\begin{center}
  { $ \begin{tabular}
{lllllllllllllllllllllllll} \hline $\scriptstyle{n\setminus k}$

             &13    &14       &15        &16       &17       &18     &19       &20     &21    &22   &23  &24   \\
\hline
14           &1     &         &          &         &         &       &         &       &      &     &    &\\
15           &2     &2        &          &         &         &       &         &       &      &     &    &\\
16           &3     &2        &1         &         &         &       &         &       &      &     &    &\\
17           &3-4   &3        &2         &2        &         &       &         &       &      &     &    &\\
18           &4     &3        &3         &2        &1        &       &         &       &      &     &    & \\
19           &5     &4        &3         &3        &2        &2      &         &       &      &     &    &  \\
20           &5-6   &5        &4         &3        &2        &2      &1        &1       &     &     &    &     \\
21           &6     &5-6      &5         &4        &3         &2       &2        &2     &1     &      &    &   \\
22           &6-7   &6        &5-6       &4-5      &4         &3      &2        &2     &2     &1     &    &         \\
23           &6-8   &6-7      &6         &5-6      &4-5       &4      &3        &2     &2     &2     &1    &     \\
24           &7-9   &6-8      &6-7       &6        &5-6       &4-5    &4        &3     &2     &2     &2     &1     \\
25           &7-9   &7-9      &6-8       &6-7      &6         &5-6    &4-5      &4     &3      &2    &2     &2         \\
\hline
\end{tabular}$
}
\end{center}
$\\$

\section*{Acknowledgements}
This work is supported by the National Natural Science Foundation of
China under Grant No.11801564.

\bibliographystyle{ieeetr}
\bibliography{bib}

\end{document}